\documentclass[multphys,vecphys]{svmult}

\usepackage{makeidx}         % allows index generation
\usepackage{graphicx}        % standard LaTeX graphics tool
\usepackage{multicol}        % used for the two-column index
\usepackage[bottom]{footmisc}% places footnotes at page bottom
\makeindex             % used for the subject index
\begin{document}

\title*{The role of tidal interactions in driving galaxy evolution.}
\author{Josefa P\'erez\inst{1,2}\and  Patricia B. Tissera\inst{2,3}\and Diego G. Lambas\inst{3,4} \and Cecilia Scannapieco\inst{2,3} }
\authorrunning{P\'erez J. et al.}
\institute{Facultad de Ciencias Astron\'omicas y Geof\'{\i}sicas, UNLP, Argentina
\texttt{jperez@fcaglp.unlp.edu.ar}
\and  Instituto de Astronom\'{\i}a y F\'{\i}sica del Espacio, Argentina \and CONICET  \and Observatorio Astron\'omico de C\'ordoba, Argentina. }

\maketitle

\begin{abstract}
We carry out a statistical analysis of galaxy pairs selected from
chemical hydrodynamical simulations with the aim at assessing the
capability of hierarchical scenarios to reproduce recent observational
results for galaxies in pairs. Particularly,
we analyse the effects of mergers and interactions on the
star formation (SF) activity, the global mean chemical properties 
and the colour distribution of  interacting galaxies. 
We also assess the effects of spurious pairs.
%Finally,in order to establish a more suitable comparation with observations,
%we also estimate possible projections effects
%on the observational results and derive 2D and 3D correlations between
%proximity to a companion and SF activity.
\end{abstract}

%Recent observations show that interactions can induce the star formation (SF)
%activity in galaxies producing an enhancement of the SF levels higher
%that those found in galaxies without a near companion (Barton et al 2000; 
%Lambas et al 2003). This is a fact that has been theoretically explained
%by, e.g. Mihos and Herquinst (1996) through numerical simulations 
%of pre-prepared mergers. They show that interactions of axisymmetrical 
%systems (without bulge or with an small one) produce a dynamical unstability
%which induces a gas inflow responsible for triggering the starbursts.
In order to determine the effects of interactions on the SF activity, 
we have used a cosmological $\Lambda$-CDM 
simulation (${\rm\Lambda}=0.7$, ${\rm\Omega}=0.3$, 
${\rm H_{0}}$=100 h $\mathrm{km \,\,s^{-1}Mpc^{-1}}$, h=0.7) run with the chemical GADGET-2 code 
(Scannapieco et al. 2005) which includes the enrichment of 
the interstellar medium by SNIa and SNII.
In agreement with observations (Lambas et al. 2003), our
results indicate that close encounters (with relative projected separation
($r_{p}$) smaller than 25 $\mathrm{kpc\,\, h^{-1}}$) can enhance the SF
activity to levels  higher than  those observed
for galaxies without a close companion (or control sample).
We also find that the triggering of SF activity 
by interactions depend on the internal properties of galaxies,
described by the depth  of their potential wells 
and gas reservoirs (P\'erez et al. 2005).
Currently passive star forming pairs (systems with SF levels
lower than that measured for the control sample) are older and more 
evolved objects, with deeper potential 
wells and less leftover gas than active star-forming systems.

%In order to assess the possible effect that projection and interlopers
%might produce in observational samples, we have constructed and analysed
%a 2D and 3D simulated catalog of galaxy pairs. 
We constructed a 3D and a projected 2D galaxy pair catalogs to study
the effects of spurious pairs.
Consistently with observations
(Mamon et al. 1986), we estimated a $\sim 27 \%$ of contamination by 
projection for systems with $r_{p}<100 \,\mathrm{kpc \,\, h^{-1}}$,
and a $\sim 19 \%$ of spurious pairs for close
systems ($r_{p}<25 \,\mathrm{kpc \,\, h^{-1}}$). 
However, comparing the 2D and 3D simulated galaxy pair samples,
we find that these levels of contamination by projection
do not affect significantly the correlations between
the SF activity and the relative separation, colour distributions 
and global chemical properties.

The analysis of  colours for galaxy pairs shows a clear 
bimodal distribution (Balogh et al. 2004) with a blue peak 
populated basically by the closest pairs with a currently strong
or recently past SF activity induced by interactions. 
Instead, the red peak is more consistent with currently
passive star forming systems, older and more evolved than
those of the blue peak. The analysis of merging and interacting
pairs shows that the former contribute with a 
larger fraction of stellar mass to the blue colours than the latter, 
demostrating the role of interactions in driving the colour bimodality.
%On the contrary, the control sample
%fits an unimodal red colour distribution, probably as a consequence
%of the efficiency of our chemical code (which not
%includes a self-consistent supernovae feedback) to transform gas into
%stars.

Finally, we study the global mean chemical abundance
of the stellar populations (SP) and the interstellar medium (ISM)
 of galaxies in pairs. 
%Galaxies were divided according to their SF
% activity in  active and passive star-forming.
%, using the mean birthrate of the control sample to segregate them.
The analysis of the chemical abundance as a function of the projected separation shows
that while the SP  are enriched with respect to galaxies without a close companion, 
regardless of their current SF activity or relative separation,
the ISM stores record fossils of the interactions. 
Galaxies with a close companion but passively forming stars
show a clear correlation of their ISM chemical abundances with distance, 
fossils of previous past interactions. Conversely, galaxies in pairs 
with active SF show an enhancement of their ISM abundances as expected.  
We also analyse the luminosity-metallicity  and mass-metallicity 
relations in order to determine if interactions 
modify the observed relations (Kobulnicky et al. 2003; Tremonti et al. 2004).
Our results show the same trends for galaxy pairs and the control sample,
indicating that these correlations might not be strongly affected by 
interactions. This point will be addressed in more detail by 
P\'erez et al (2006, in preparation).

%\section{Section Heading}
%\label{sec:1}

% Always give a unique label
% and use \ref{<label>} for cross-references
% and \cite{<label>} for bibliographic references
% use \sectionmark{}
% to alter or adjust the section heading in the running head
%Your text goes here. Use the \LaTeX\ automatism for your citations
%\cite{monograph}.

%\subsection{Subsection Heading}
%\label{sec:2}
%Your text goes here.

%\subsubsection{Subsubsection Heading}
%Your text goes here. Use the \LaTeX\ automatism for cross-references as
%well as for your citations, see Sect.~\ref{sec:1}.

%\paragraph{Paragraph Heading} %
%Your text goes here.

%\subparagraph{Subparagraph Heading.} Your text goes here.%
%
%\index{paragraph}
% Use the \index{} command to code your index words
%

\subsection*{Acknowledgements.}
MJP thanks the LOC of this conference for their financial support.
This work was partially funded by Fundaci\'on Antorchas, CONICET and LENAC.

%%%%%%%%%%%%%%%%%%%%%%%% referenc.tex %%%%%%%%%%%%%%%%%%%%%%%%%%%%%%
% sample references
% "physics"
%
% Use this file as a template for your own input.
%
%%%%%%%%%%%%%%%%%%%%%%%% Springer-Verlag %%%%%%%%%%%%%%%%%%%%%%%%%%

%
% BibTeX users please use
% \bibliographystyle{}
% \bibliography{}

\begin{thebibliography}{99.}
% Use the following syntax and markup for your references
%
% Monographs



\bibitem{journal} Balogh M. L., Eke V., Miller C., et al., 2004, MNRAS 348, 1355

\bibitem{journal} Kobulnicky H. A.\& Kewley L. J., 2004, ApJ 617, 240
\bibitem{journal} Lambas, D. G., Tissera, P. B., Alonso, M. S. Coldwell, G. 2003,
MNRAS 346, 1189

\bibitem{journal} Mamon G.A., 1986, ApJ, 307, 436

\bibitem{journal} Perez J., Tissera, P. B., Lambas, D. G.\& Scannapieco C., 2005, 
A\&A accepted (astro-ph/0510327)

\bibitem{journal} Scannapieco C., Tissera P.B., White S.D.M. \& Springel V., 2005, MNRAS 364, 552



\end{thebibliography}

%%%%%%%%%%%%%%%%%%%%%%%%%%%%%%%%%%%%%%%%%%%%%%%%%%%%%%%%%%%%%%%%%%%%%%  }

%%%%%%%%%%%%%%%%%%%%%%%%%%%%%%%%%%%%%%%%%%%%%%%%%%%%%%%%%%%%%%%%%%%%%%

\printindex
\end{document}